\documentclass[12pt]{iopart}

\usepackage{graphicx,bbm}

\def\tr{\,{\rm tr}\,}
\def\ket#1{|#1\rangle}
\def\bra#1{\langle#1|}

\def\ave#1{\langle #1 \rangle}
\def\avec#1{\langle #1 \rangle_{\rm c}}
\def\ii{{\rm i}}

\def\tit#1{{\em #1},}
\def\etal#1{#1}

\begin{document}

\title{A matrix product solution for a nonequilibrium steady state of an XX chain}

\author{Marko \v Znidari\v c}
\address{Department of Physics, Faculty of Mathematics and Physics,
  University of Ljubljana, Ljubljana, Slovenia}
\date{\today}

\begin{abstract}
A one dimensional XX spin chain of finite length coupled to reservoirs at both ends is solved exactly in terms of a matrix product state ansatz. An explicit reprsentation of matrices of fixed dimension $4$ independent of the chain length is found. Expectations of all observables are evaluated, showing that all connected correlations, apart from the nearest neighbor $z-z$, are zero.
\end{abstract}


\section{Introduction}

Simplification leading to understanding of essential features of physical systems is one of the leading principles in theoretical physics. Finding simple solutions to seemingly complicated models is one of the ways to approach this goal. In quantum physics the complexity of a system grows exponentially with the number of particles. Even if an exact solution is possible representing it in a compact way is nontrivial for many-body systems. One approach to represent a given state is to express its expansion coefficients in a suitable basis in terms of product of matrices, the so-called matrix product states~\cite{Fannes:92,Klumper:91}, used before in a wholly different context of 2d classical lattice model~\cite{Hakim83}. If quantum state is only weakly correlated the resulting matrices can be small. In condensed matter such an ansatz has been used to exactly describe ground states of many low dimensional spin systems, among the first one for instance the AKLT chain~\cite{AKLT} or ladder system~\cite{Kolezhuk:97}. In addition to ground states, description in terms of products of matrices is also successfully used in algorithms for simulation of quantum systems~\cite{Murg}. Crucial for the efficiency of such simulation is the necessary matrix dimension $D$. Unfortunately, generic coherent quantum evolution will cause $D$ to grow exponentially with the simulation time~\cite{pre:07}, rendering simulation inefficient. However, in some special cases of Heisenberg evolution of certain operators in integrable systems, like transverse Ising chain, the dimension $D$ is small and does not grow with time~\cite{pre:07,Iztok}. Efficient matrix product description with time-independent $D$ is possible also for certain open versions of the same integrable models~\cite{clark:10}.

Besides quantum systems, matrix product states are also widely used in classical stochastic models. There a matrix product formulation is used to study nonequilibrium stochastic lattice models, in particular their nonequilibrium stationary state (NESS). Since its first use~\cite{Derrida93} for the exact solution of a one-dimensional asymmetric exclusion process~\cite{asep} a matrix product formulation has been utilized in a plethora of models, for a review see~\cite{exclusion}. Expectations in the NESS can sometimes be calculated with the help of algebra only, sometimes though an explicit representation of matrices is required. It is not known in general when a finite dimensional matrix representation is possible.

In the present work we provide an explicit finite dimensional matrix product solution for a stationary state of quantum system in nonequilibrium situation. This extends the applicability of matrix product states in quantum domain from ground states and time evolutions to the NESS. The model we consider is a one-dimensional XX spin chain coupled to reservoirs at chain ends. Using the Jordan-Wigner transformation one can reformulate the system in terms of spinless fermions. In the fermionic language the system is composed of free fermions. Hamiltonian part is therefore trivially integrable. In fact, for our choice of bath Lindblad operators the whole superoperator ${\cal L}$ is quadratic in fermionic operators and can therefore be diagonalized in the operator space~\cite{3rd}. An explicit solution for the first two orders in the driving can be also obtained as a particular case of an analytic solution for the XX chain with dephasing~\cite{JSTAT}. Slightly different model resulting in the same solution, in which XX chain repeatedly interacts with independent spins of the bath, has been explicitly solved in~\cite{Karevski:09}. For studies of nonequilibrium states in a doubly infinite chain see~\cite{pillet,ogata,platini07}. Our primary goal here is therefore not to solve the system but instead to provide a compact form for the exact NESS in terms of a matrix product ansatz with the matrices of low dimension $D=4$ that is independent of the chain length. This in turn enables one to explicitly evaluate the expectation value in the NESS of an arbitrary observable. 

Hamiltonian of the XX spin chain is given by
\begin{equation}
H=\sum_{j=1}^{n-1} (\sigma_j^{\rm x} \sigma_{j+1}^{\rm x} +\sigma_j^{\rm y} \sigma_{j+1}^{\rm y}),
\label{eq:H}  
\end{equation}
with standard Pauli matrices; lower indices running from $j=1,\ldots,n$ denote a site position. Dynamics of the spin chain coupled to the environment will be described in an effective way using the Lindblad master equation~\cite{lindblad}
\begin{equation}
\frac{{\rm d}}{{\rm d}t}{\rho}=\ii [ \rho,H ]+ {\cal L}^{\rm bath}_{\rm L}(\rho)+{\cal L}^{\rm bath}_{\rm R}(\rho)={\cal L}(\rho).
\label{eq:Lin}
\end{equation}
To induce a nonequilibrium situation we couple the chain at the first and the last site to a ``bath'', modelled here by linear operator ${\cal L}^{\rm bath}$ which is expressed in terms of Lindblad operators $L^{\rm L}_{1,2}$ acting on the first site, and $L^{\rm R}_{1,2}$ acting on the last site,
\begin{equation}
{\cal L}^{\rm bath}_{\rm L,R}(\rho)=\sum_k \left( [ L_k^{\rm L,R} \rho,L_k^{{\rm L,R}\dagger} ]+[ L_k^{\rm L,R},\rho L_k^{{\rm L,R}\dagger} ] \right).
\end{equation}
We take the simplest Lindblad operators of the form
\begin{equation}
L^{\rm L}_1=\sqrt{\Gamma_{\rm L}(1-\frac{\mu}{2}+\bar{\mu})}\,\sigma^+_1,\qquad L^{\rm L}_2=\sqrt{\Gamma_{\rm L}(1+\frac{\mu}{2}-\bar{\mu})}\, \sigma^-_1,
\end{equation}
on the $1$st site, while on the $n$th site we have
\begin{equation}
L^{\rm R}_1=\sqrt{\Gamma_{\rm R}(1+\frac{\mu}{2}+\bar{\mu})}\,\sigma^+_n,\qquad L^{\rm R}_2=\sqrt{\Gamma_{\rm R}(1-\frac{\mu}{2}-\bar{\mu})}\, \sigma^-_n,
\end{equation}
$\sigma^\pm_j=(\sigma^{\rm x}_j \pm {\rm i}\, \sigma^{\rm y}_j)/2$. Such bath operators induce an imbalance in magnetization, causing a flow of magnetization from one to the other end. For bath operators only, the stationary state, for which ${\cal L}^{\rm bath}_{\rm L}(\rho)=0$ holds, is diagonal in the eigenbasis of $\sigma^{\rm z}_1$, with the expectation value $\ave{\sigma_1^{\rm z}}=\tr{(\rho\, \sigma_1^{\rm z})}=\bar{\mu}-\frac{\mu}{2}$, and correspondingly for ${\cal L}^{\rm bath}_{\rm R}$. Therefore, parameter $\bar{\mu}$ plays the role of average magnetization, $\mu$ its difference between the right and left ends, while $\Gamma_{\rm L,R}$ are coupling strengths. Matrix representation of superoperators ${\cal L}^{\rm bath}_{\rm L,R}$ can be found in the~\ref{sec:appendix}. We are interested in a stationary solution of the whole master equation, ${\cal L}(\rho)=0$, which we call a nonequilibrium stationary state, NESS for short, and denote simply by $\rho$.

\section{Matrix product ansatz}
We shall write the NESS $\rho$ with the matrix product operator (MPO) ansatz,
\begin{equation}
\rho=\frac{1}{2^n}\sum_{\alpha_1,\alpha_2,\ldots,\alpha_n} \bra{1} A_1^{(\alpha_1)} A_2^{(\alpha_2)} \cdots A_n^{(\alpha_n)}\ket{1}\,\, \sigma_1^{\alpha_1} \sigma_2^{\alpha_2} \cdots \sigma_n^{\alpha_n}.
\end{equation}
Indices $\alpha_i$ run over labels of Pauli matrices forming an operator basis, $\alpha_i \in \{{\rm x,y,z},\mathbbm{1} \}$, with the convention $\sigma^{\mathbbm{1}}_j=\mathbbm{1}_j$, matrices $A_i^{(\alpha_i)}$ are of dimension $D \times D$, while $\ket{1}$ is a $D$-dimensional unit vector. We arbitrarily choose its components to be $\delta_{j,1}$. 

The goal is to write the NESS in terms of as small matrices as possible. Our method of solution shall be the following. In Ref.~\cite{JSTAT}, an exact solution for an XX model with dephasing has been provided of which our current XX chain is a special limit (the limit of zero dephasing). It has been observed that in a model without dephasing there are no long-range correlations, that is, all connected correlations are zero. This leads us to think that one could perhaps construct an exact matrix product operator solution just by observing nontrivial one and two-point observables; in our case these are magnetization, current and $z-z$ correlations. Therefore, based on the solution from~\cite{JSTAT}, we are first going to construct an MPO that accounts only for few-point observables in the NESS. Then we are going to show that such MPO in fact provides an exact soluton, that is, all observables and not just few-point are reproduced correctly. To show this we are going to use the algebraic properties of the matrices found.

In~\cite{JSTAT} it has been found that (for $\Gamma_{\rm L}=\Gamma_{\rm R}=1$, $\bar{\mu}=0$) the solution is (upto normalization)
\begin{equation}
\fl
\rho \approx \mathbbm{1}+\frac{\mu}{4}(-\sigma_1^{\rm z}+\sigma_n^{\rm z})-\frac{\mu}{4}\sum_{j=1}^{n-1} (\sigma_j^{\rm x} \sigma_{j+1}^{\rm y}-\sigma_j^{\rm y} \sigma_{j+1}^{\rm x})-\frac{\mu^2}{16}(\sigma_1^{\rm z} \sigma_n^{\rm z}+\sum_{j=1}^{n-1} \sigma_j^{\rm z} \sigma_{j+1}^{\rm z})+ {\cal O}(\mu^2).
\label{eq:rho}
\end{equation}
In the above expression we write only linear terms in $\mu$ and $z-z$ term of order $\mu^2$. All other are inessential for the following discussion\footnote{Two quadratic terms not written in eq.(\ref{eq:rho}) are $\frac{\mu^2}{32}(-\sigma_1^{\rm z}+\sigma_n^{\rm z})\sum_{k=1}^{n-1}(\sigma_k^{\rm x} \sigma_{k+1}^{\rm y}-\sigma_k^{\rm y} \sigma_{k+1}^{\rm x})+\frac{\mu^2}{32}\sum_{k=1}^{n-1}(\sigma_k^{\rm x} \sigma_{k+1}^{\rm y}-\sigma_k^{\rm y} \sigma_{k+1}^{\rm x})(-\sigma_1^{\rm z}+\sigma_n^{\rm z})$ and $\frac{\mu^2}{32}\sum_{k\neq l=1}^{n-1}(\sigma_k^{\rm x} \sigma_{k+1}^{\rm y}-\sigma_k^{\rm y} \sigma_{k+1}^{\rm x})(\sigma_l^{\rm x} \sigma_{l+1}^{\rm y}-\sigma_l^{\rm y} \sigma_{l+1}^{\rm x})$.}. How can we write such state in terms of products of matrices? As an easy overture, we start with a simpler operator $\rho$ obtained by keeping in the NESS (\ref{eq:rho}) only terms with $\sigma_j^{\rm z}$, i.e., dropping the current term in eq.(\ref{eq:rho}). One can easily convince oneself that the matrices $A_j^{(\rm z)}=a_j \ket{1}\bra{1}+ \ket{2}\bra{1} - \frac{\mu^2}{16} \ket{1}\bra{2}$ and $A_j^{(\mathbbm{1})}=\ket{1}\bra{1}$, where $a_1=-\frac{\mu}{4}$, $a_n=\frac{\mu}{4}$, $a_{2,\ldots,n-1}=a_{\rm bulk}=0$, while trivially $A_j^{(\rm x,y)}=0$, gives the wanted state. Therefore, for this simple operator MPO ansatz with dimension $D=2$ would suffice. To describe also the current though, we have to allow for at least two additional basis states in matrices $A_i$. We found that matrices of size $D=4$ are sufficient. To correctly describe all terms of order $\mu$ and $\mu^2$ (also those of order $\mu^2$ that are not explicitly written out in eq.(\ref{eq:rho})) the following set of matrices gives the correct operator,
\begin{eqnarray}
A^{(z)}_i&=
\left( \begin{array}{cccc}
a_i & -t^2 & 0 & 0 \\
1 & 0 & 0 & 0 \\
0 & 0 & 0 & 0 \\
0 & 0 & 0 & 0\\
\end{array} \right), \qquad \quad
A^{(\mathbbm{1})}_i&=
\left( \begin{array}{cccc}
1 & 0 & 0 & 0 \\
0 & 0 & 0 & 0 \\
0 & 0 & 0 & t \\
0 & 0 & t & 0\\
\end{array} \right),\nonumber \\
A^{(x)}&=(-P,-P,P,P,-P,\ldots),\quad A^{(y)}&=(-R,R,R,-R,-R,\ldots), \nonumber \\
P&=
\left( \begin{array}{cccc}
0 & 0 & 0 & t \\
0 & 0 & 0 & 0 \\
1 & 0 & 0 & 0 \\
0 & 0 & 0 & 0\\
\end{array} \right), \qquad \qquad
R&=
\left( \begin{array}{cccc}
0 & 0 & t & 0 \\
0 & 0 & 0 & 0 \\
0 & 0 & 0 & 0 \\
1 & 0 & 0 & 0\\
\end{array} \right),
\label{eq:MPO} 
\end{eqnarray}
where we use the notation $A^{(\rm x,y)}=(A_1^{(\rm x,y)},A_2^{(\rm x,y)},\ldots)$. Matrices $A^{(\rm x)}_i$ and $A^{(\rm y)}_i$ are periodic with period 4, $A^{(\rm x)}_{i+4}=A^{(\rm x)}_i$, and can be concisely written as $A^{(\rm x)}_j=(\cos{\frac{\pi}{2}j}-\sin{\frac{\pi}{2}j})\,P$, while $A^{(\rm y)}_j=-(\cos{\frac{\pi}{2}j}+\sin{\frac{\pi}{2}j})\,R$. We shall show that the MPO with the above matrices (\ref{eq:MPO}) and appropriately chosen three parameters $t$, $a_1$, $a_n$ and $a_{\rm bulk}=a_{2,\ldots,n-1}$ is an exact NESS solution for the Lindblad equation (\ref{eq:Lin}) with arbitrary $\Gamma_{\rm L,R},\mu,\bar{\mu}$. To show this we are going to use an algebraic approach similar to the one used in solutions of classical stochastic processes.

Let us denote by ${\cal L}_{i,i+1}^{\rm (H)}$ the superoperator from the commutator part of master equation corresponding to the nearest-neighbor 2-spin term in the Hamiltonian, $\sigma_i^{\rm x} \sigma_{i+1}^{\rm x} +\sigma_i^{\rm y} \sigma_{i+1}^{\rm y}$ in our case. Writing four MPO matrices in a vector of matrices as $A_i=(A^{(\rm x)}_i,A^{(\rm y)}_i,A^{(\rm z)}_i,A^{(\mathbbm{1})}_i)$, we can form all $16$ different products of two matrices at consecutive sites through $A_i \otimes A_{i+1}$ (a $4 \times 4$ matrix, each element being a product of two $D \times D$ matrices). Depending on the action of ${\cal L}_{i,i+1}^{\rm (H)}$ on products of operators, i.e., on ${\cal L}_{i,i+1}^{\rm (H)}(A_i \otimes A_{i+1})$, see also the~\ref{sec:appendix}, solving for NESS is relatively simple in two cases: (i) if the NESS is separable or has only two-particle entanglement, like for instance in valence bond states, then we can have ${\cal L}_{i,i+1}^{\rm (H)}(A_i \otimes A_{i+1})=0$; (ii) other relatively simple situation is when ${\cal L}_{i,i+1}^{\rm (H)}(A_i \otimes A_{i+1})$ results in a divergence-like term, that is 
\begin{equation}
{\cal L}_{i,i+1}^{\rm (H)}(A_i \otimes A_{i+1})=A_i \otimes M_{i+1} - M_i \otimes A_{i+1},
\label{eq:bulk}
\end{equation}
with some matrices $M_i=(M_i^{(\rm x)},M_i^{(\rm y)},M_i^{(\rm z)},M_i^{(\mathbbm{1})})$. Note that this ansatz is a trivial inhomogeneous extension of a standard procedure used in classical nonequilibrium systems~\cite{Hinrichsen}. The reason to allow for spatially dependent matrices is that we want to find the MPO solution with the smallest $D$. We find that representation with $D=4$ is possible irrespective of the chain length $n$~\footnote{For an explicit small dimensional MPO construction of some simple operators see~\cite{Frowis}.}. Any inhomogeneous solution can be written as a site-independent one but with larger matrices. One possibility is to make block site-independent matrix $\tilde{A}$ of size $D\cdot n$ out of site-dependent matrices $A_i$ as
\begin{equation}
\tilde{A}=\left( \begin{array}{ccccc}
0 & A_1 & 0 & \cdots & 0 \\
0 & 0 & A_2 & \cdots & 0 \\
\vdots & \vdots & \vdots & \ddots & \vdots \\
0 & 0 & 0 & \cdots & A_{n-1} \\
A_n & 0 & 0 &\cdots & 0\\
\end{array} \right).
\end{equation}
If equation (\ref{eq:bulk}) holds terms from consecutive ${\cal L}_{i,i+1}^{\rm (H)}$ will pairwise cancel, except at the boundaries. To ensure stationarity we have to enforce an additional condition at the boundaries. For our case of bath acting on a single boundary spin we get two equations,
\begin{equation}
\langle 1 | \left[ {\cal L}_{\rm L}^{\rm bath}(A_1)-M_1\right]=0,\quad \left[ {\cal L}_{\rm R}^{\rm bath}(A_n)+M_n \right]|1 \rangle =0.
\label{eq:rob}
\end{equation}
If one manages to find such $A_i$ and the associated $M_i$ that eqs.(\ref{eq:bulk}) and (\ref{eq:rob}) are satisfied one has found the NESS solution of the master equation (\ref{eq:Lin}). The condition (\ref{eq:bulk}) can in our case of the Pauli basis of the operator space be written as a set of $4^2$ matrix equations of size $D \times D$. As we have only $2\cdot 4$ matrices $A_i$ and $M_i$ of size $D \times D$, with $8D^2$ unknown parameters, it is not guaranteed that the solution exists.

We are now going to show that for the XX model it actually does exist. We are going to find an explicit representation of matrices $M_i$, showing that they, together with $A_i$ (\ref{eq:MPO}), satisfy equations (\ref{eq:bulk}) and (\ref{eq:rob}). Similarly as for $A_i^{(\rm x,y)}$, matrices $M_i^{(\rm x,y)}$ also have the periodicity 4. Their explicit ($D=4$)-dimensional representation in the bulk, that is for the sites $i=2,\ldots,n-1$, is
\begin{eqnarray}
M^{(\rm \mathbbm{1})}_i&=-2 a_{\rm bulk}\left( \begin{array}{cccc}
0 & 0 & 0 & 0 \\
0 & 0 & 0 & 0 \\
0 & 0 & 0 & 1 \\
0 & 0 & 1 & 0\\
\end{array} \right), \quad M^{(\rm z)}_i&=-2\left( \begin{array}{cccc}
2t & 0 & 0 & 0 \\
0 & 2t & 0 & 0 \\
0 & 0 & 0 & 1 \\
0 & 0 & 1 & 0\\
\end{array} \right), \nonumber \\
M^{\rm (x)}&=(S,S,-S,-S,S,\ldots),\qquad M^{\rm (y)}&=(T,-T,-T,T,T,\ldots), \nonumber \\
S&=2\left( \begin{array}{cccc}
0 & 0 & 0 & 0 \\
0 & 0 & 0 & -1 \\
0 & t & 0 & 0 \\
0 & 0 & 0 & 0\\
\end{array} \right), \qquad \quad
T&=2\left( \begin{array}{cccc}
0 & 0 & 0 & 0 \\
0 & 0 & -1 & 0 \\
0 & 0 & 0 & 0 \\
0 & t & 0 & 0\\
\end{array} \right).
\label{eq:mbar}
\end{eqnarray}
One can convince oneself by direct calculation that the above $M_i$ together with $A_i$ (\ref{eq:MPO}) satisfy condition (\ref{eq:bulk}), written out in full the~\ref{sec:appendix}, eq.~(\ref{eq:eqs}). To also fulfill the two boundary conditions (\ref{eq:rob}), $M_1$ and $M_n$ must have additional matrix elements and parameters $t,a_1,a_n$ and $a_{\rm bulk}$ must take specific values. We have $M^{\rm (x)}_1=S+2(a_{\rm bulk}-a_1) | 1 \rangle \langle 4|$, $M^{\rm (y)}_1=T+2(a_{\rm bulk}-a_1) | 1 \rangle \langle 3|$ and  $M^{\rm (z)}_1=M^{\rm (z)}_2+4t(a_{\rm bulk}-a_1) | 1 \rangle \langle 2|$. On the right end we have $M^{\rm (z)}_n=M^{\rm (z)}_2-\frac{4}{t}(a_{\rm bulk}-a_n) | 2 \rangle \langle 1|$, $M^{\rm (x)}_n={\rm sgnx}_n (S+\frac{2}{t}(a_{\rm bulk}-a_n) | 3 \rangle \langle 1|)$, $M^{\rm (y)}_n={\rm sgny}_n (T+\frac{2}{t}(a_{\rm bulk}-a_n) | 4 \rangle \langle 1|)$, where ${\rm sgnx}_n$ and ${\rm sgny}_n$ are the signs in front of last $S$ or $T$ in eq.(\ref{eq:mbar}) and depend on the site $n$. In addition, for the boundary terms to be zero the values of parameters must be
\begin{eqnarray}
t &=& \mu \frac{\Gamma_{\rm L} \Gamma_{\rm R}}{(1+\Gamma_{\rm L} \Gamma_{\rm R})(\Gamma_{\rm L}+\Gamma_{\rm R})}, \nonumber \\
a_1 &=&\bar{\mu}-\frac{\mu}{2}\, \frac{(\Gamma_{\rm L}-\Gamma_{\rm R})+\Gamma_{\rm L}\Gamma_{\rm R}(\Gamma_{\rm L}+\Gamma_{\rm R})}{(1+\Gamma_{\rm L} \Gamma_{\rm R})(\Gamma_{\rm L}+\Gamma_{\rm R})} \nonumber \\
a_{2,\ldots,n-1} &=& a_{\rm bulk}=\bar{\mu}-\frac{\mu}{2}\, \frac{(\Gamma_{\rm L}-\Gamma_{\rm R})(1-\Gamma_{\rm L} \Gamma_{\rm R})}{(1+\Gamma_{\rm L} \Gamma_{\rm R})(\Gamma_{\rm L}+\Gamma_{\rm R})} \nonumber \\
a_n &=&\bar{\mu}-\frac{\mu}{2}\, \frac{(\Gamma_{\rm L}-\Gamma_{\rm R})-\Gamma_{\rm L}\Gamma_{\rm R}(\Gamma_{\rm L}+\Gamma_{\rm R})}{(1+\Gamma_{\rm L} \Gamma_{\rm R})(\Gamma_{\rm L}+\Gamma_{\rm R})}.
\label{eq:a}
\end{eqnarray}
Note that we have $a_1-a_{\rm bulk}=-t \Gamma_{\rm L}$, $a_n-a_{\rm bulk}=t \Gamma_{\rm R}$ and $a_n-a_1=t(\Gamma_{\rm L}+\Gamma_{\rm R})$. Parameters (\ref{eq:a}) together with matrices (\ref{eq:MPO}) form an exact MPO solution of the NESS for the XX chain.

If we add a homogeneous magnetic field in the $z$-direction to our Hamiltonian, that is the term of the form $B\sum_{i=1}^n \sigma_i^{\rm z}$, the model can again be solved exactly. In fact, the single-site superoperator due to the magnetic field acts as ${\cal L}_i^{\rm (B)}(\sigma_i^{\rm x})= 2B\, \sigma_i^{\rm y}$, ${\cal L}_i^{\rm (B)}(\sigma_i^{\rm y})=- 2B\,\sigma_i^{\rm x}$, while ${\cal L}_i^{\rm (B)}(\sigma_i^{\rm z,\mathbbm{1}})= 0$. Due to the symmetry of the NESS without the field (see also the explicit form of all nonzero terms given in~\cite{JSTAT}) and the minus sign in the action of ${\cal L}_i^{\rm (B)}$, we have $\sum_i {\cal L}_i^{\rm (B)}(\rho)=0$. This means that the solution presented (\ref{eq:MPO}) is also the exact NESS in the presence of an arbitrary homogeneous field of strength $B$. 

\section{Expectations of observables}

With an explicit representation of matrix product solution at hand we can evaluate expectations of various operators in the NESS. Let us first evaluate expectation of a series of $\sigma_i^{\rm z}$ operators, like $\ave{\sigma_i^{\rm z} \sigma_j^{\rm z} \cdots}$. One or two $\sigma^{\rm z}$s are simple to evaluate by direct calculation, we obtain $\ave{\sigma_i^{\rm z}}=a_i$ and $\ave{\sigma_i^{\rm z} \sigma_j^{\rm z}}=a_i a_j-t^2 \delta_{i+1,j}$. Connected correlation function of $n$ operators, $\avec{ O_{i_1} O_{i_2} \cdots O_{i_n}}$, is obtained from an ordinary expectation value by subtracting product of all connected correlations where each involves less than $n$ operators. For instance, $\avec{O_i}=\ave{O_i}$, $\avec{O_i O_j}=\ave{O_i O_j}-\avec{O_i}\avec{O_j}$, $\avec{ O_{i_1} O_{i_2} O_{i_3}}=\ave{ O_{i_1} O_{i_2} O_{i_3}}-\avec{O_{i_1}}\avec{O_{i_2} O_{i_3}}-\avec{O_{i_2}}\avec{O_{i_1} O_{i_3}}-\avec{O_{i_3}}\avec{O_{i_1} O_{i_2}}-\avec{O_{i_1}}\avec{O_{i_2}}\avec{O_{i_3}}$, and so on. Connected correlation function of two $\sigma^{\rm z}$s is therefore nonzero only on neighboring sites, $\avec{\sigma_i^{\rm z} \sigma_j^{\rm z}}=-t^2 \delta_{i+1,j}$. We are now going to show by induction that all higher order connected correlation functions involving more than two $\sigma^{\rm z}$s are identically zero. Assume that the statement holds for products of upto $n$ $\sigma^{\rm z}$s. Denoting by $Z=\sigma_i^{\rm z} \sigma_j^{\rm z} \cdots \sigma_k^{\rm z}=\tilde{Z}\sigma_k^{\rm z}$ a product of $n$ not-necessarily neighboring $\sigma^{\rm z}_i$, we would like to show that the connected correlation $\avec{Z \sigma_l^{\rm z}}$ is zero for $n \ge 2$. Assuming connected correlations of more than two $\sigma^{\rm z}$s are zero, we can write
\begin{eqnarray}
\avec{Z \sigma_l^{\rm z}}&=&\ave{Z \sigma_l^{\rm z}}-\left({\sum}'\avec{Z} \right) \avec{\sigma_l^{\rm z}}-\left({\sum}' \avec{\tilde{Z}}\right)\avec{\sigma_k^{\rm z} \sigma_l^{\rm z}}= \nonumber \\
&=& \ave{Z \sigma_l^{\rm z}}-\ave{Z}\avec{\sigma_l^{\rm z}}-\ave{\tilde{Z}}\avec{\sigma_k^{\rm z} \sigma_l^{\rm z}},
\label{eq:highz}
\end{eqnarray}
where we denoted by ${\sum}' \avec{Z}$ a sum of all products of connected correlations, each involving $\le n$ operators $\sigma^{\rm z}$, for instance, ${\sum}' \avec{\sigma_i^{\rm z} \sigma_j^{\rm z} }= \avec{\sigma_i^{\rm z} \sigma_j^{\rm z}}+\avec{\sigma_i^{\rm z}}\avec{\sigma_j^{\rm z}}$. We used the fact that to have a nonzero connected correlation $\sigma_l^{\rm z}$ must be paired with at most one other $\sigma^{\rm z}_j$ and that it must be its neighbor (non nearest-neighbor connected correlations are zero). Using our explicit MPO representation (\ref{eq:MPO}) of matrices for $A^{\rm (z)}$ and $A^{(\mathbbm{1})}$ we will now show that the right-hand side of (\ref{eq:highz}) is zero. First, observe that the matrix corresponding to the product of matrices occurring in the operator $Z$ has an upper-left $2\times 2$ block equal to
\begin{equation}
Z=\left( \begin{array}{cc}
\ave{Z} & -t^2 \ave{\tilde{Z}} \\
\, * & * \\
\end{array}
\right).
\end{equation}
Upper left element is $\ave{Z}$ by definition while the upper right follows through a simple multiplication of matrix corresponding to $\tilde{Z}$ by $A^{\rm (z)}_k$. With the explicit form of $Z$ we have
\begin{equation}
\ave{Z \sigma_l^{\rm z}}=\ave{Z}\avec{\sigma_l^{\rm z}}-t^2 \ave{\tilde{Z}}\,\delta_{k+1,l},
\end{equation}
where the Kronecker delta takes into account multiplication of the matrix for $Z$ by $A^{(\mathbbm{1})}_{k+1}$ in the case of non-neighboring sites $k$ and $l$. Plugging this into eq.(\ref{eq:highz}) we see that the $(n+1)$-point connected correlation of $\sigma^{\rm z}$ is indeed zero. This completes the proof.

Single point expectations $\ave{\sigma_i^{\rm x}}$ and $\ave{\sigma_i^{\rm y}}$ are zero. Among two-point expectations on neighboring sites, besides $z-z$, the only nonzero terms are two from the spin current, that is $\ave{\sigma_i^{\rm x} \sigma_{i+1}^{\rm y}}=-\ave{\sigma_i^{\rm y} \sigma_{i+1}^{\rm x}}=-t$. Therefore, 
the expectation value of the spin current operator $j_k=2(\sigma_k^{\rm x}\sigma_{k+1}^{\rm y}-\sigma_k^{\rm y}\sigma_{k+1}^{\rm x})$ is $\ave{j_k}=-4t$. If we have a product of non-overlapping current operators at different sites, $\ave{j_i j_k j_l \cdots}$ (with $k>i+1, l>k+1,\ldots$), the expectation value is simple. Observe that $A^{\rm (x)}_i A^{\rm (y)}_{i+1}=-t(\ket{1}\bra{1}+\ket{3}\bra{3})$. Because of the form of $A^{(\mathbbm{1})}$, if we have a product of only $A^{(\mathbbm{1})}_j$s and $p$ terms $A^{\rm (x)}_i A^{\rm (y)}_{i+1}$ at various sites $i$, their expectation value is simply equal to $t^p$. This means that the connected correlation function of non-overlapping operators $j_i$ is zero, $\avec{j_i j_k j_l \cdots}=0$, apart from $\avec{j_i}=-4t$. If two current operators overlap, i.e. $j_i j_{i+1}$, with the hermitean $(j_i j_{i+1}+j_{i+1} j_{i})/2=-4(\sigma_i^{\rm x}\sigma_{i+2}^{\rm x}+ \sigma_i^{\rm y}\sigma_{i+2}^{\rm y})$, the corresponding product of matrices is $A_i^{\rm (x)}A_{i+1}^{(\mathbbm{1})}A_{i+2}^{\rm (x)}=-(t^2 \ket{1}\bra{1}+t\ket{3}\bra{4})$. From this form, similarly as before, we can see that for products of $j_i$ the term $\ket{3}\bra{4}$ is not important, resulting again in all connected correlations being zero. Similar argument holds for products of more than two overlapping current operators.

Finally, let us discuss expectations of products of $\sigma^{\rm z}_i$ and $j_k$, the only remaining nonzero terms in the NESS. For overlapping sites, $j_i \sigma_i^{\rm z}+\sigma_i^{\rm z} j_i=0$ (the same for $j_{i+1}$), and we have to consider only non-overlapping operators. Two-point expectation is $\ave{j_i \sigma_j^{\rm z}}=-4t a_j$, connected correlation being therefore $0$. Again, taking into account that if we consider only products of $A^{(\mathbbm{1})}_j$, $A^{\rm (z)}_k$ and $A^{\rm (x)}_i A^{\rm (y)}_{i+1}$, the term $\ket{3}\bra{3}$ in $A^{\rm (x)}_i A^{\rm (y)}_{i+1}$ is irrelevant because there are no $\ket{3}\bra{1}$, $\ket{4}\bra{1}$ or their hermitean conjugates in any of the three matrices. The expectations are therefore trivial products of scalar quantities in front of $\ket{1}\bra{1}$ terms, resulting in all connected correlations being zero. If we have a product of neighboring currents, like for instance in $(j_i j_{i+1}+j_{i+1} j_{i})/2$, similar argument holds. Incidentally, we also see that the expectation of the energy current $j^{\rm E}_i=2(\sigma_{i-1}^{\rm x}\sigma_{i}^{\rm z}\sigma_{i+1}^{\rm y}- \sigma_{i-1}^{\rm y}\sigma_{i}^{\rm z}\sigma_{i+1}^{\rm x})$, such that ${\rm i}[\sigma_{i}^{\rm x}\sigma_{i+1}^{\rm x}+\sigma_{i}^{\rm y}\sigma_{i+1}^{\rm y},H]=j_i^{\rm E}-j_{i+1}^{\rm E}$, is zero. In the NESS with our choice of baths no energy current flows. 

Another way to write the NESS is to rewrite it in terms of an exponential function as $\rho=\exp{(-\tilde{H})}$. Doing the calculation we observe that the operator $\tilde{H}$ contains only few nonzero terms. These are $\sigma_i^{\rm z}$ at all $n$ sites (with different prefactors on different sites), spin current $j_k=2b^{(2)}_k$ at all $n-1$ sites (the same prefactor on all sites) as well as $h^{(2k+1)}_i$, with $k=1,\ldots$, and $b^{(2k)}_i$ with $k=2,\ldots$, where $h_j^{(k)}=\sigma_j^{\rm x} \sigma_{j+1}^{\rm z} \cdots \sigma_{j+k-2}^{\rm z} \sigma_{j+k-1}^{\rm x}+\sigma_j^{\rm y} \sigma_{j+1}^{\rm z} \cdots \sigma_{j+k-2}^{\rm z} \sigma_{j+k-1}^{\rm y}$ and $b_j^{(k)}=\sigma_j^{\rm x} \sigma_{j+1}^{\rm z} \cdots \sigma_{j+k-2}^{\rm z} \sigma_{j+k-1}^{\rm y}-\sigma_j^{\rm y} \sigma_{j+1}^{\rm z} \cdots \sigma_{j+k-2}^{\rm z} \sigma_{j+k-1}^{\rm x}$. This shows that the NESS in the open XX chain can not be exactly written as a quasi-equilibrium generalized Gibbs state $\rho \sim \exp{(\sum_j \kappa_j Q_j)}$ with locally varying fields $\kappa_j$, and $Q_j$ conserved quantities of the corresponding Hamiltonian system~\cite{Jaynes}. Namely, conserved quantities in integrable systems can depend on small changes in the system, for instance on boundary conditions. For 1d XX chain with periodic boundary conditions two infinite sequences of conserved quantities exist, one set are $Q_{2k}^+=\sum_j h^{(2k)}_j$ and $Q_{2k+1}^+=\sum_j b^{(2k+1)}_j$, another $Q_{2k}^-=\sum_j b^{(2k)}_j$ and $Q_{2k+1}^-=\sum_j h^{(2k+1)}_j$. If we change boundary conditions to open, only half of conserved quantities survive, that is, for open boundary conditions conserved quantities are only $\tilde{Q}_{2k}=Q^+_{2k}+q_{2k}$ and $\tilde{Q}_{2k+1}=Q^-_{2k+1}+q_{2k+1}$, where $q_j$ are additional boundary terms~\cite{Grabowski}. First few conserved quantities for open XX chain are $\tilde{Q}_2=\sum_{j=1}^{n-1} h^{(2)}_j$, $\tilde{Q}_3=\sum_{j=1}^{n-2}h^{(3)}_j \, +\sigma_1^{\rm z}+\sigma_n^{\rm z}$, $\tilde{Q}_4=\sum_{j=1}^{n-3} h^{(4)}_j \, - h^{(2)}_1-h^{(2)}_{n-1}$, $\tilde{Q}_5=\sum_{j=1}^{n-4} h^{(5)}_j \, + \sigma_2^{\rm z}-h^{(3)}_1+\sigma^{\rm z}_{n-1}-h^{(3)}_{n-2}$. We can see that $\tilde{H}$ can not be written as a sum of conserved quantities $\tilde{Q}$ of an open chain neither that of a periodic one. A change in the boundary condition can therefore globally influence constants of motion of Hamiltonian system. This sensitivity also translates to open systems described by master equation: a change in boundary operators of the bath can have a large influence on the NESS. In short, for open integrable systems there is an ambiguity which ``conserved'' quantities $Q_j$ should one use in the generalized Gibbs ensemble, or, in other words, the $Q_j$ depend on the bath. Such non-universality, where the functional form of NESS does not depend only on the Hamiltonian but also on reservoirs, is probably generic situation for open versions of integrable systems.

To summarize, we have shown that all connected correlations, apart from neighboring $\avec{\sigma_i^{\rm z} \sigma_{i+1}^{\rm z}}$, are identically zero. This should not come as a surprise. In fact, working in fermionic language the system is quadratic, therefore all expectations of products of fermionic operators can be evaluated in terms of two-point expectations using Wick's theorem.

\section{Conclusion}
We have found an exact solution of an open XX chain in terms of the matrix product ansatz. An explicit ($D=4$)-dimensional representation of matrices is found, enabling to evaluate arbitrary expectations. All connected correlation functions, apart from the correlation function of magnetization at nearest-neighbor sites, are identically zero. The results presented extend the applicability of matrix product states to quantum nonequilibrium systems. Furthermore, the method of solution is an algebraic one borrowed from the field of classical stochastic processes where it has been used very successfully. It is hopped that this will lead to new exactly solvable nonequilibrium quantum systems. One such instance is an open XX chain with dephasing, a solvable diffusive model~\cite{JSTAT} where, based on our experience, a compact matrix product solution is also possible. Unfortunately though, in an even more interesting XXZ model the algebra seems to be more complicated.

\ack 
Support by the Program P1-0044 and the Grant J1-2208 of the Slovenian Research Agency is acknowledged.

\appendix

\section{}
\label{sec:appendix}

\subsection{Representation of bath superoperators}
Using a basis of Pauli matrices, and tensor products thereof, a single-site superoperator for the bath is
\begin{equation}
{\cal L}^{\rm bath}_{\rm L}=\Gamma_{\rm L}
\left( \begin{array}{cccc}
-2 & 0 & 0 & 0 \\
0 & -2 & 0 & 0 \\
0 & 0 & -4 & -2(\mu-2\bar{\mu}) \\
0 & 0 & 0 & 0\\
\end{array} \right).
\end{equation}
Superoperator ${\cal L}^{\rm bath}_{\rm R}$ for the right bath is obtained by replacing $\Gamma_{\rm L}$ with $\Gamma_{\rm R}$ and $\mu$ by $-\mu$.
Basis elements are ordered as $(\sigma^{\rm x},\sigma^{\rm y},\sigma^{\rm z},\mathbbm{1})$. Matrix representation gives us the operation of ${\cal L}^{\rm bath}$ on the expansion coefficients of operators in the Pauli basis, e.g., writing $\rho=\sum c_\alpha \sigma^\alpha$, we have ${\cal L}\rho=\rho'=\sum c'_\alpha \sigma^\alpha$, with $c'_\alpha=\sum_\beta {\cal L}_{\alpha,\beta}\, c_\beta$. Alternatively, the same matrices can be thought of as transforming matrices in the MPO ansatz, e.g., $[{\cal L}^{\rm bath}_{\rm L}(A_1)]_i=\sum_{j=1}^{4}[{\cal L}^{\rm bath}_{\rm L}]_{i,j} [A_1]_j$.

\subsection{Representation of ${\cal L}_{i,i+1}^{\rm (H)}$}

Superoperator ${\cal L}_{i,i+1}^{\rm (H)}$ transforms operators according to their commutator with the Hamiltonian. It is therefore fully specified by its operation on a $16$ dimensional basis of 2-site operators. We choose products of Pauli matrices as a basis. As an example, for the XX chain we have for instance, ${\cal L}_{i,i+1}^{\rm (H)}(\mathbbm{1}_i \sigma_{i+1}^{\rm x})=-2 \sigma_i^{\rm y} \sigma_{i+1}^{\rm z}$. Acting on MPO ansatz, this for instance gives ${\cal L}_{i,i+1}^{\rm (H)}( \langle \cdots A^{(\mathbbm{1})}_i A^{\rm (x)}_{i+1} \cdots \rangle \, \mathbbm{1}_i \sigma_{i+1}^{\rm x} )= \langle \cdots (-2 A^{(\mathbbm{1})}_i A^{\rm (x)}_{i+1}) \cdots \rangle \, \sigma_i^{\rm y} \sigma_{i+1}^{\rm z}$. In the transformed state, the term that comes in the product of matrices at the position of $A^{\rm (y)}_i A^{\rm (z)}_{i+1}$ is therefore equal to $(-2 A^{(\mathbbm{1})}_i A^{\rm (x)}_{i+1})$. Because we are interested in how ${\cal L}_{i,i+1}^{\rm (H)}$ transforms MPO, it is handy to represent its action in terms of a $4 \times 4$ matrix whose element $[{\cal L}_{i,i+1}^{\rm (H)}]_{\alpha,\beta}$ gives us the term in front of the $\sigma_i^\alpha \sigma_{i+1}^\beta$. For the XX chain we have 
\begin{equation}
{\cal L}_{i,i+1}^{\rm (H)}=2 \left( \begin{array}{cccc}
0 & B & A^{(\mathbbm{1})}_i A^{\rm (y)}_{i+1} & A^{\rm (z)}_i A^{\rm (y)}_{i+1} \\
-B & 0 & -A^{(\mathbbm{1})}_i A^{\rm (x)}_{i+1} & - A^{\rm (z)}_i A^{\rm (x)}_{i+1} \\
A^{\rm (y)}_i A^{(\mathbbm{1})}_{i+1} &  -A^{\rm (x)}_i A^{(\mathbbm{1})}_{i+1} & 0 &  -C \\
 A^{\rm (y)}_i A^{\rm (z)}_{i+1} & -  A^{\rm (x)}_i A^{\rm (z)}_{i+1}&  C & 0\\
\end{array} \right),
\end{equation}
where $B=A^{\rm (z)}_i A^{(\mathbbm{1})}_{i+1}-A^{(\mathbbm{1})}_i A^{\rm (z)}_{i+1}$ and $C=A^{\rm (x)}_i A^{\rm (y)}_{i+1}- A^{\rm (y)}_i A^{\rm (x)}_{i+1}$, explicitly giving the transformation
\begin{equation}
{\cal L}_{i,i+1}^{\rm (H)}(\rho) =\sum_{\alpha_i,\alpha_{i+1}} \langle \cdots ([{\cal L}_{i,i+1}^{\rm (H)}]_{\alpha_i,\alpha_{i+1}}) \cdots \rangle\, \sigma_i^{\alpha_i} \sigma^{\alpha_{i+1}}_{i+1}.
\end{equation}

Written explicitly, the algebra (\ref{eq:bulk}) of matrices $A_j$ and $M_j$ in the bulk that is induced by ${\cal L}_{i,i+1}^{\rm (H)}$ is for the XX model
\begin{equation}
\begin{array}{rcl}
2 (A^{\rm (z)}_i A^{(\mathbbm{1})}_{i+1}-A^{(\mathbbm{1})}_i A^{\rm (z)}_{i+1}) &=& A^{\rm (x)}_i M^{\rm (y)}_{i+1}-M_i^{\rm (x)} A_{i+1}^{\rm (y)}  \\
-2 (A^{\rm (z)}_i A^{(\mathbbm{1})}_{i+1}-A^{(\mathbbm{1})}_i A^{\rm (z)}_{i+1}) &=& A^{\rm (y)}_i M^{\rm (x)}_{i+1}-M_i^{\rm (y)} A_{i+1}^{\rm (x)}\\
2 (A^{\rm (x)}_i A^{\rm (y)}_{i+1}-A^{\rm (y)}_i A^{\rm (x)}_{i+1}) &=& A^{(\mathbbm{1})}_i M^{\rm (z)}_{i+1}-M_i^{(\mathbbm{1})} A_{i+1}^{\rm (z)}  \\
-2 (A^{\rm (x)}_i A^{\rm (y)}_{i+1}-A^{\rm (y)}_i A^{\rm (x)}_{i+1}) &=& A^{\rm (z)}_i M^{(\mathbbm{1})}_{i+1}-M_i^{\rm (z)} A_{i+1}^{(\mathbbm{1})}  \\
0 &=& A^{\rm (x)}_i M^{\rm (x)}_{i+1}-M_i^{\rm (x)} A_{i+1}^{\rm (x)}  \\
0 &=& A^{\rm (y)}_i M^{\rm (y)}_{i+1}-M_i^{\rm (y)} A_{i+1}^{\rm (y)}  \\
0 &=& A^{\rm (z)}_i M^{\rm (z)}_{i+1}-M_i^{\rm (z)} A_{i+1}^{\rm (z)}  \\
0 &=& A^{(\mathbbm{1})}_i M^{(\mathbbm{1})}_{i+1}-M_i^{(\mathbbm{1})} A_{i+1}^{(\mathbbm{1})}  \\
2 A_i^{(\mathbbm{1})} A_{i+1}^{\rm (y)} &=& A^{\rm (x)}_i M^{\rm (z)}_{i+1}-M_i^{\rm (x)} A_{i+1}^{\rm (z)}  \\
2 A_i^{\rm (z)} A_{i+1}^{\rm (y)} &=& A^{\rm (x)}_i M^{(\mathbbm{1})}_{i+1}-M_i^{\rm (x)} A_{i+1}^{(\mathbbm{1})}  \\
-2 A_i^{(\mathbbm{1})} A_{i+1}^{\rm (x)} &=& A^{\rm (y)}_i M^{\rm (z)}_{i+1}-M_i^{\rm (y)} A_{i+1}^{\rm (z)}  \\
-2 A_i^{\rm (z)} A_{i+1}^{\rm (x)} &=& A^{\rm (y)}_i M^{(\mathbbm{1})}_{i+1}-M_i^{\rm (y)} A_{i+1}^{(\mathbbm{1})}  \\
2 A_i^{\rm (y)} A_{i+1}^{(\mathbbm{1})} &=& A^{\rm (z)}_i M^{\rm (x)}_{i+1}-M_i^{\rm (z)} A_{i+1}^{\rm (x)}  \\
-2 A_i^{\rm (x)} A_{i+1}^{(\mathbbm{1})} &=& A^{\rm (z)}_i M^{\rm (y)}_{i+1}-M_i^{\rm (z)} A_{i+1}^{\rm (y)}  \\
2 A_i^{\rm (y)} A_{i+1}^{\rm (z)} &=& A^{(\mathbbm{1})}_i M^{\rm (x)}_{i+1}-M_i^{(\mathbbm{1})} A_{i+1}^{\rm (x)}  \\
-2 A_i^{\rm (x)} A_{i+1}^{\rm (z)} &=& A^{(\mathbbm{1})}_i M^{\rm (y)}_{i+1}-M_i^{(\mathbbm{1})} A_{i+1}^{\rm (y)}
\end{array}
\label{eq:eqs}
\end{equation}
Representation given in eq.(\ref{eq:MPO}) and (\ref{eq:mbar}) satisfies this algebra for the arbitrary values of four parameters $a_1,a_n,a_{\rm bulk}$ and $t$.

\end{document}